\documentclass{article}
\usepackage{ijcai22}

\usepackage{times}
\usepackage{soul}
\usepackage{url}
\usepackage[hidelinks]{hyperref}
\usepackage[utf8]{inputenc}
\usepackage[small]{caption}
\usepackage{graphicx}
\usepackage{amsmath}
\usepackage{amsthm}
\usepackage{booktabs}
\usepackage{algorithm}
\usepackage{algorithmic}
\usepackage[switch]{lineno}
\usepackage{multirow}
\usepackage{enumitem}
\usepackage{amsfonts,amsmath}

\newtheorem{theorem}{Theorem}

\title{Learning Adaptable Risk-Sensitive Policies to Coordinate in Multi-Agent General-Sum Games}

\author{
Ziyi Liu$^1$
\and
Yongchun Fang$^2$
\affiliations
NanKai University
\emails
$^1$liuzy@mail.nankai.edu.cn,
$^2$fangyc@nankai.edu.cn
}

\begin{document}

\maketitle

\begin{abstract}
In general-sum games, the interaction of self-interested learning agents commonly leads to socially worse outcomes, such as defect-defect in the iterated stag hunt (ISH). Previous works address this challenge by sharing rewards or shaping their opponents’ learning process, which require too strong assumptions. In this paper, we demonstrate that agents trained to optimize expected returns are more likely to choose a safe action that leads to guaranteed but lower rewards. However, there typically exists a risky action that leads to higher rewards in the long run only if agents cooperate, e.g., cooperate-cooperate in ISH. To overcome this, we propose using action value distribution to characterize the decision's risk and corresponding potential payoffs. Specifically, we present Adaptable Risk-Sensitive Policy (ARSP). ARSP learns the distributions over agent's return and estimates a dynamic risk-seeking bonus to discover risky coordination strategies. Furthermore, to avoid overfitting training opponents, ARSP learns an auxiliary opponent modeling task to infer opponents' types and dynamically alter corresponding strategies during execution. Empirically, agents trained via ARSP can achieve stable coordination during training without accessing opponent's rewards or learning process, and can adapt to non-cooperative opponents during execution. To the best of our knowledge, it is the first method to learn coordination strategies between agents both in iterated prisoner's dilemma (IPD) and iterated stag hunt (ISH) without shaping opponents or rewards, and can adapt to opponents with distinct strategies during execution. Furthermore, we show that ARSP can be scaled to high-dimensional settings.

\end{abstract}

\section{Introduction}

Most existing works in multi-agent reinforcement learning (MARL) focus on the fully cooperative ~\cite{coma,qmix,qtran,qplex,rmix} and competitive ~\cite{go,grandmaster} settings. However, these settings only represent a fraction of potential real-world multi-agent environments. General-sum games, in which multiple self-interested learning agents optimize their own rewards independently and win-win outcomes are only possible through coordination~\cite{general-sum}, describe many domains such as self-driving cars~\cite{self-drive} and human-robot interactions ~\cite{human-robot}.

Coordination is often coupled with risk. There are many scenarios where there exists a safe action that leads to guaranteed but socially and individually lower rewards, and a risky action that leads to higher rewards only if agents cooperate, such as alleviating traffic congestion~\cite{self-driving}. This kind of multi-agent coordination problem presents unique challenges that are not presented in single-agent learning~\cite{lola,wang2021emergent}, and simply applying MARL algorithms to train self-interested agents typically converge on unconditional mutual defection, which is the globally worst outcome~\cite{rpg,lu2022model}.


To avoid such catastrophic outcomes, one set of approaches use explicit reward shaping to force agents to be prosocial, such as by making agents care about rewards of their partners~\cite{prosocial,rpg}, which can be viewed as shaping the risk degree of coordination strategies. However, this requires the strong assumption that agents involved are altruistic and can access other agents' reward function. Other works either treat partners as stationary~\cite{liam,wang2018towards} or take into account their learning step in order to shape their policy~\cite{lola}. By contrast, we focus on general settings where multiple decentralized, separately-controlled, and partially-observable agents interact in the environment and only care about maximizing their own rewards - while the objective is still to increase the probability of coordination. Furthermore, policies learned during training should be adaptable so that the agent is able to dynamically alter its strategies between different modes, e.g., either cooperate or defect, w.r.t. its test-time opponent’s behavior.

In this paper, one key insight is that learning from opponent's history behaviors allows the agent to adapt to different opponents during execution. Moreover, given that the other learning agents are non-stationary, decision-making over the agent's return distributions enables the agent to model uncertainties resulting from other agents' behaviors and alter its risk preference, i.e., from risk-neutral to risk-seeking, to discover coordination strategies. Motivated by the analysis above, we propose ARSP, an Adaptable Risk-Sensitive MARL algorithm and our contributions are summarized as follows:

\textbf{Leading to stable coordination in decentralized general-sum games.} We estimate a dynamic risk-seeking bonus to encourage agents to discover risky coordination strategies. Specifically, the risk-seeking bonus is estimated using a complete distortion risk measure Wang's Transform (WT)~\cite{wang2000class} and only affects the action selection procedure instead of shaping environment rewards, and decreases throughout training which leading to an unbiased policy. 

\textbf{Adaptable to different opponents during execution.} Policies learned independently can overfit other agents’ policies in the training phase, failing to adapt to different opponents during execution~\cite{lanctot2017unified}. We further propose to train each learning agent with two objectives: a standard Quantile Regression objective~\cite{qr,qrdqn} and a supervised agent modeling objective which models the behaviors of opponents and affects the intermediate representation learning of the value network. The auxiliary opponent modeling task allows the policy to be influenced by opponent's past behaviors, forcing the intermediate representation to adapt to new opponents. 

\textbf{Evaluating in multi-agent settings.} We evaluate ARSP in four different Markov games:
Iterated Stag Hunt (ISH)~\cite{wang2021emergent,rpg}, Iterated Prisoners’ Dilemma (IPD)~\cite{lu2022model,lola,wang2018towards},
Monster-Hunt~\cite{rpg,zhou2021continuously} and  Escalation~\cite{rpg,prosocial}. Compared with baseline methods, ARSP agents learn substantially faster, achieves stable coordination during training and can adapt to non-cooperative opponents during execution. Furthermore, we show that our method can be scaled to high-dimensional settings.

\section{Related Work}
\textbf{Risk-sensitive and distributional RL.} Risk-sensitive policies, which depend upon more than mean of the outcomes, enable agents to handle the intrinsic uncertainties arising from the stochasticity of the environment. In MARL, the intrinsic uncertainties are amplified due to the non-stationarity and partial observability created by other agents that change their policies during the learning procedure~\cite{wang2022influencing,papoudakis2019dealing,hernandez2017survey}. Distributional RL~\cite{bellemare2017distributional,qrdqn} provides a new perspective for optimizing policy under different risk preferences within a unified framework~\cite{iqn,markowitz2021risk}. With distributions of return, it is able to approximate value function under different risk measures, such as Conditional Value at Risk (CVaR)~\cite{cvar,chow2015risk} and WT~\cite{wang2000class}, and thus produces risk-averse or risk-seeking policies. Qiu et al.~\cite{rmix} propose RMIX with the CVaR measure as risk-averse policies. Similar ideas are proposed in LH-IQN~\cite{lyu2018likelihood} and DFAC~\cite{sun2021dfac}. But unlike these works, which focus on the fully cooperative settings and leverage distributional RL to alleviate stochasticity or generate risk-averse policies, our method utilizes return distributions to quantify the decision risk in general-sum games and yields a novel risk-seeking exploration bonus to encourage agents to achieve stable coordination.

\textbf{Test-time adaptation across different opponents.}
Many real world scenarios require agents to adapt to different opponents during execution. However, most of existing works focus on learning a fixed and team-dependent policy in fully cooperative settings~\cite{vdn,qmix,qtran,rmix,qplex}, which can not adapt to slightly altered environments or different opponents during execution. Other works either use the population-based training method to train an adaptable agent~\cite{rpg,lupu2021trajectory,strouse2021collaborating}, or adapt to different opponents under the Tit-for-Tat principle~\cite{wang2018towards,peysakhovich2017consequentialist} in the IPD. Our work is closely related to test-time training methods~\cite{sun2020test,hansen2020self}. However, they focus on image recognition or single agent policy adaption. Ad hoc teamwork~\cite{stone2010ad,zhang2020multi} also requires agents to adapt to new teams, but they focus on cooperative games and has different concerns with us.

\textbf{Opponent modeling.} Our approach to learning adaptable policies can be viewed as a kind of opponent modeling method~\cite{albrecht2018autonomous}. Most existing works either focus on modeling opponent's intention~\cite{wang2013probabilistic,raileanu2018modeling} or exploit opponent learning dynamics~\cite{lola,zhang2010multi}. We consider a more general setting where agents can not only model opponents during training but can transfer their learned knowledge to different opponents at test-time.
Policy reconstruction methods~\cite{raileanu2018modeling} which make explicit predictions about opponents' actions are similar with our approach. However, instead of predicting the opponent's future actions and converging to a fixed policy, ARSP learns from opponents' past behaviors to infer their strategies and can transfer its learned knowledge to different opponents during execution.

\section{Preliminaries}
\textbf{Stochastic games.}
In this work, we consider multiple self-interested learning agents interact with each other. We model the problem as a Partially-Observable Stochastic Game (POSG)~\cite{shapley1953stochastic,hansen2004dynamicposg}, which consists of $N$ agents, a state space $\mathcal{S}$ describing the possible configurations of all agents, a set of actions $\mathcal{A}^{1},\ldots,\mathcal{A}^{N}$ and a set of observations $\mathcal{O}^{1},\ldots,\mathcal{O}^{N}$ for each agent. At each time step, each agent $i$ receives its own observation ${o}^{i}\in \mathcal{O}^{i}$, and selects an action $a^i\in \mathcal{A}^i$ based on a stochastic policy $\pi^{i}:\mathcal{O}^i\times \mathcal{A}^i \mapsto [0,1]$, which results in a joint action vector $\boldsymbol{a}$. The environment then produces a new state $s^{\prime}$ based on the transition function $P(s^{\prime}|s,\boldsymbol{a})$. Each agent $i$ obtains rewards as a function of the state and its action $R^i:\mathcal{S}\times\mathcal{A}^i\mapsto \mathbb{R}$. The initial states are determined by the distribution $\rho:\mathcal{S}\mapsto [0,1]$. We treat the reward "function" $R^i$ of each agent as a random variable to emphasize its stochasticity, and use $Z^{\pi^{i}}(s,a^i)=\sum_{t=0}^{T}\gamma^{t}R^{i}(s_t,a_t^i)$ to denote the random variable of the cumulative discounted rewards where $S_0=s$, $A^{i}_0=a^i$,$\gamma$ is a discount factor and $T$ is the time horizon.

\textbf{Distorted expectation.} Distorted expectation is a risk weighted expectation of value distribution under a specific distortion function~\cite{wirch2001distortion}. A function $g:[0,1]\mapsto[0,1]$ is a distortion function if it is non-decreasing and satisfies $g(0)=0$ and $g(1)=1$~\cite{balbas2009properties}. The distorted expectation of $Z$ under $g$ is defined as $\Psi(Z)=\int_{0}^{1} F_{Z}^{-1}(\tau) d g(\tau)=\int_{0}^{1} {g}^{\prime}(\tau) F_{Z}^{-1}(\tau) d \tau$, where $F_{Z}^{-1}$ is the quantile function at $\tau\in[0,1]$ for the random variable $Z$. We introduce two common distortion functions as follows:

\begin{itemize}[leftmargin=*]
    \item \textbf{CVaR} is the expectation of the lower or upper tail of the value distribution, corresponding to risk-averse or risk-seeking policy respectively. Its distortion function is $g(\tau)=\min(\tau/\alpha, 1)$ (risk-averse) or $ \max\left(0,1-(1-\tau)/\alpha\right)$ (risk-seeking), $\alpha\in (0,1)$ denotes confidence level.
    
    \item \textbf{WT} is proposed by Wang~\cite{wang2000class}: $g_{\lambda}(\tau)=\Phi\left(\Phi^{-1}(\tau)+\lambda\right),$ where $\Phi$ is the distribution of a standard normal. The parameter $\lambda$ is called the market price of risk and reflects systematic risk. $\lambda>0$ for risk-averse and $\lambda<0$ for risk-seeking.
\end{itemize}
$\text{CVaR}_{\alpha}$ assigns a 0-value to all percentiles below the $\alpha$ or above $1-\alpha$ significance level which leads to erroneous decisions in some cases~\cite{balbas2009properties}. Instead, WT is a complete distortion risk measure and ensures using all the information in the original loss distribution which makes training much more stable, and we will empirically demonstrate it in sec. \ref{experiment}.

\section{Methods}
In this section, we describe our proposed Adaptable Risk-Sensitive Policy (ARSP) method. We first introduce the motivation behind our risk-seeking bonus in sec. \ref{example} and then clarify the mathematical formulation of the bonus in sec. \ref{risk-seeking}. Furthermore, we propose the auxiliary opponent modeling task to learn adaptable policies in sec. \ref{aom}. The details of test-time policy adaptation under different opponents are present in sec. \ref{adaptation}.


\subsection{Motivation}
\label{example}

Let's consider a classical matrix game in the game theory: Stag Hunt (SH). Two players choose either Stag action or Hare action at the same time. If both agents choose to hunt stag, they receive the highest payoff $a$. If they both hunt hare, each of them will receive d reward. However, if one player choose to hunt stag and the other hunt hare, the player who hunt stag will receive the lowest payoff c or even be punished, and the player who hunt hare will receive a guaranteed reward b. Table \ref{tab:sh} models this situation. There exists two pure strategy Nash Equilibrium (NE): (Stag, Stag) and (Hare, Hare). The Stag NE is the only pareto optimal NE, but it is risky since if the other agent defects, the agent will receive a big loss c, e.g., $c = -10$. The following theorem similar to ~\cite{rpg} shows if agents compute the expected payoff function's gradient to update its policy parameters , then the probability the strategies converge to the Stag NE via policy gradient is very low.

\begin{table}[h]
\centering
\caption{The stag-hunt game, $a > b \geq d > c$.}
\label{tab:sh}
\begin{tabular}{c|c|c|}
     & Stag & Hare \\ \hline
Stag & a, a & c, b \\ \hline
Hare & b, c & d, d \\ \hline
\end{tabular}
\end{table}

\begin{theorem}
\label{them1}
Suppose $a - b = \epsilon (d - c)$ for some $0 < \epsilon < 1$ and each play has its own policy $\pi_{i}(\theta_i)$. $P[\pi_{i}(\theta_i) = S] = \theta_i$ and $P[\pi_{i}(\theta_i) = H] = 1 - \theta_i$. If initialize $\theta_{1}, \theta_{2}\sim$ Unif $[0, 1]$, then the probability that agents discover the pareto optimal NE via policy gradient is $\frac{\epsilon^2}{\epsilon^2+2\epsilon+1}$ 
\end{theorem}

We provide the proof in the appendix. Theorem \ref{them1} shows that if the risk is high, i.e., c is low, then the $\epsilon$ is low and the probability of converging to Stag NE via policy gradient is very low. It is noteworthy that the expectation ignores the complete information of agent's payoff distribution, e.g., the upper and lower tail information when the return distribution is asymmetric. 

\begin{figure}[h]
  \centering
  \includegraphics[width=0.85\linewidth]{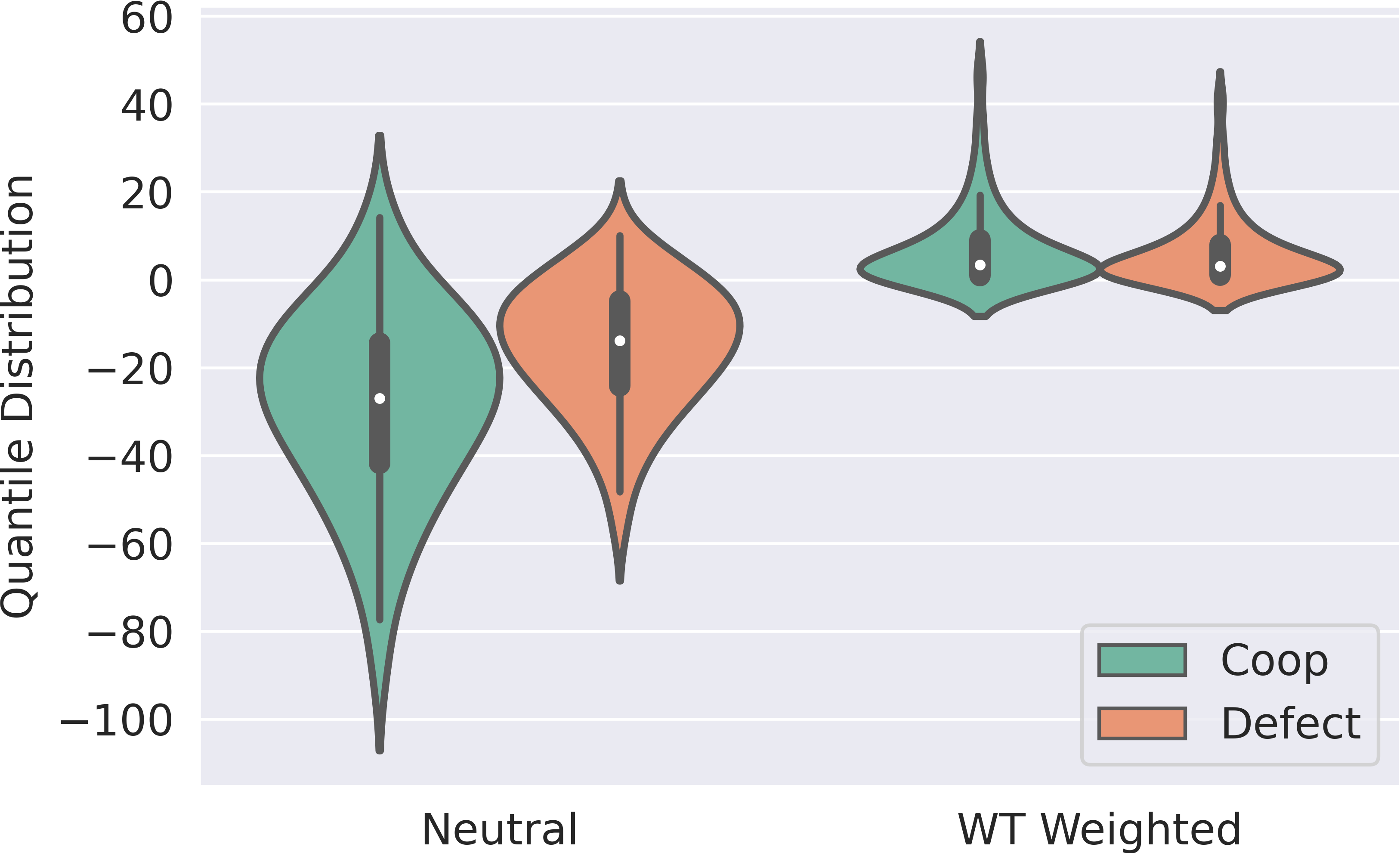}
  \caption{Quantile value distribution of cooperation and defection in Iterated Stag Hunt weighted by WT compared with risk-neutral policy.}
  \label{fig:1}
\end{figure}

Fig.\ref{fig:1} left part shows the quantile value distribution of hunting stag (cooperation, Coop) and hunting hare (defection, Defect) in the Iterated Stag Hunt (ISH) learned by a risk-neutral policy. We repeat the stag hunt matrix game ten times in the ISH. The return expectation of Defect is higher than that of Coop, but the Coop distribution has a longer upper tail which means that it has a higher potential payoff. However, the expectation 
unable to express this information. Fig.\ref{fig:1} right part shows the WT weighted quantile distribution learned by a risk-seeking policy. The risk-seeking policy gives the upper tail higher weight when computing distorted expectations, so the agent is more tolerant of the risk and pays more attention to the potential payoff.

\subsection{Risk-Seeking Bonus}
\label{risk-seeking}

\begin{figure*}
    \centering
    \includegraphics[width=\linewidth]{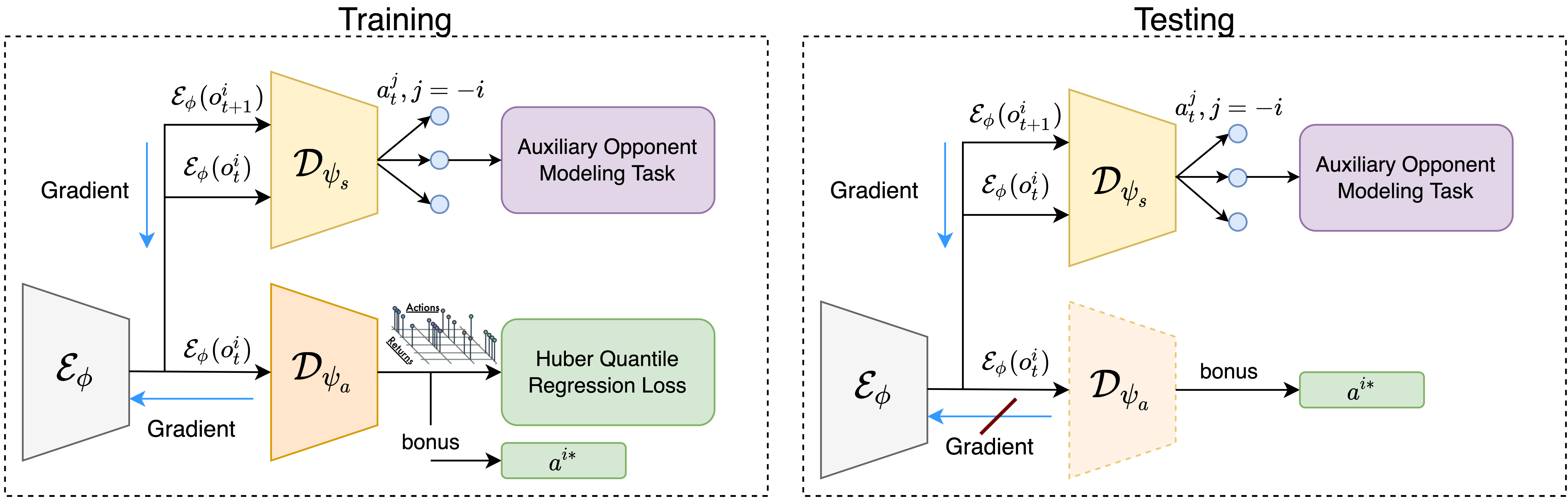}
    \caption{\textbf{Left:} Diagram of ARSP architecture during training. Outputs of $\mathcal{E}_{\phi}$ are fed into $\mathcal{D}_{\psi_{a}}$ and $\mathcal{D}_{\psi_{s}}$, so features are shared between policy and auxiliary opponent modeling. The prediction head $\mathcal{D}_{\psi_{s}}$ outputs other agents' actions. \textbf{Right:} Test-Time policy adaptation. The agent can not receive environment rewards during testing, so we only optimize the auxiliary opponent modeling objective.}
    \label{GRSP-AR}
\end{figure*}

Based on analysis above, we propose to use WT distortion function to reweight the expectation of quantile distribution. By following ~\cite{qrdqn}, we first represent the return distribution of each agent $i$ with policy $\pi^i$ by a uniform mix of $M$ supporting quantiles:
\begin{equation}
    Z_{\theta}^{\pi^i}(o^i, a^i) \doteq \frac{1}{M} \sum_{k=1}^{M} \delta_{\theta_{k}^{\pi^i}(o^i, a^i)}
\end{equation}
where $\delta_x$ denotes a Dirac Delta function at $x\in \mathbb{R}$, and each $\theta_k^{\pi^i}$ is an estimation of the quantile corresponding to the quantile fractions $\hat{\tau}_{k} \doteq \frac{\tau_{k-1}+\tau_{k}}{2}$ with $\tau_{k} \doteq \frac{k}{M}$ for $0\leq k \leq M$. The state-action value $Q^{\pi^i}(o^i, a^i)$ can then be approximated by$\frac{1}{M} \sum_{k=1}^{M}\theta_{k}^{\pi^i}(o^i, a^i)$.

Furthermore, we propose the risk-seeking bonus $\Psi$ defined as:
\begin{equation}
    \Psi(Z_{\theta}^{\pi^i})=\int_{0}^{1} {g}_{\lambda}^{\prime}(\tau) F_{Z_{\theta}^{\pi^i}}^{-1}(\tau) d \tau \approx \frac{1}{M}\sum_{k=1}^{M}{g}_{\lambda}^{\prime}(\hat{\tau}_k)\theta^{i}_{k},
\end{equation}
where $g_{\lambda}^{\prime}(\tau)$ is the derivatives of WT distortion function at $\tau\in[0,1]$, and $\lambda$ controls the risk-seeking level. Fig.\ref{fig:1} right part shows the WT weighted quantile distribution in which the upper quantile values are multiplied by bigger weights and lower quantile values are multiplied by smaller weights to encourage agents to discover risky coordination strategies.

A naive approach to exploration would be to use the variance of the estimated distribution as a bonus. ~\cite{mavrin2019distributional} shows that the exploration bonus from truncated variance outperforms bonus from the variance. Specifically, the Right Truncated Variance tells about lower tail variability and the Left Truncated Variance tells about upper tail variability. For instantiating optimism in the face of uncertainty, the upper tail variability is more relevant than the lower tail, especially if the estimated distribution is asymmetric. So we adopt the Left Truncated Variance of quantile distribution to further leverage the intrinsic uncertainty for efficient exploration. The left truncated variance is defined as

\begin{equation}
\mathcal{\sigma}_{+}^{2}=\frac{1}{2 M} \sum_{j=\frac{M}{2}}^{M}\left(\theta_{\frac{M}{2}}-\theta_{j}\right)^{2},
\end{equation}
and analysed in ~\cite{mavrin2019distributional}. The index starts from the median, i.e., $M/2$, rather than the mean due to its well-known statistical robustness ~\cite{huber2011robust,rousseeuw2011robust}. We anneal the two exploration bonuses dynamically so that in the end we produce unbiased policies. The anneal coefficients are defined as $c_{tj}=c_j \sqrt{\frac{\log t}{t}}, j=1,2$ which is the parametric uncertainty decay rate~\cite{koenker2001quantile}, and $c_j$ is a constant factor. This approach leads to choosing the action such that
\begin{equation}
a^{i*}=\arg\max_{a^i\in \mathcal{A}^i}\left( Q^{\pi^i}(o^i,a^i)+c_{t1} \Psi(Z^{\pi^i}(o^i,a^i))+c_{t2} \sqrt{\sigma_{+}^2(o^i,a^i)} \right )
\end{equation}

\begin{figure*}
    \centering
    \includegraphics[width=\linewidth]{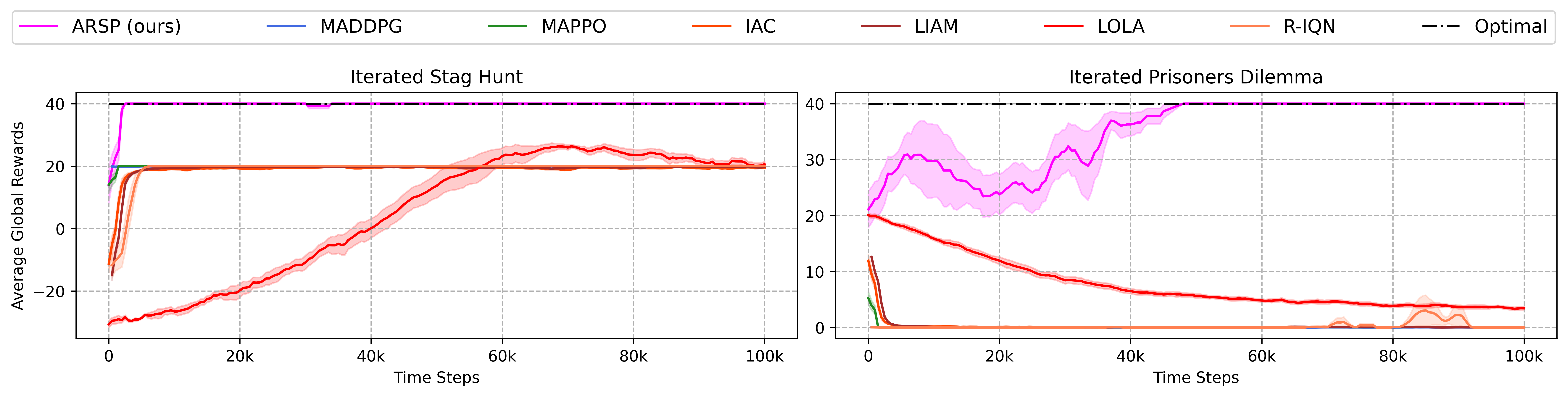}
    \caption{Mean evaluation returns for ARSP, MADDPG, MAPPO, IAC, LIAM and LOLA on two repeated matrix games. The average global rewards equal to 40 means that all agents have learned coordination strategy, i.e., cooperating at each time step.}
    \label{fig:img}
\end{figure*}

These quantile estimates are trained using the Huber~\cite{huber1992robust} quantile regression loss. The loss of the quantile value network of each agent $i$ at time step $t$ is then given by

\begin{equation}
\mathcal{J}\left(o_{t}^{i}, a_{t}^{i}, r_{t}^{i}, o_{t+1}^{i}; \theta^{i}\right)=\frac{1}{M} \sum_{k=0}^{M-1} \sum_{j=0}^{M-1} \rho_{\hat{\tau}_{k}}^{\kappa}\left(\delta_{k j}^{ti}\right)
\end{equation}

where $\delta_{k j}^{t i}\doteq r_t^i+\gamma \theta_{j}^{i}\left(o_{t+1}^{i}, \pi^{i}\left(o_{t+1}^{i}\right)\right)-\theta_{k}^{i}(o_t^i, a_t^i)$, and $\rho_{\hat{\tau}_{k}}^{\kappa}(x)\doteq \left|\hat{\tau}_{k}-\mathbb{I}\left\{x<0\right\}\right| \frac{\mathcal{L}_{\kappa}\left(x\right)}{\kappa}$ where $\mathbb{I}$ is the indicator function and $\mathcal{L}_{\kappa}(x)$ is the Huber loss:

\begin{equation}
\mathcal{L}_{\kappa}(x) \doteq \begin{cases}\frac{1}{2} x^{2} & \text { if } x \leq \kappa \\ \kappa\left(|x|-\frac{1}{2} \kappa\right) & \text { otherwise }\end{cases}
\end{equation}


\subsection{Auxiliary Opponent Modeling Task}\label{aom}

In order to alter the agent's strategies under different opponents, we share parameters between policy and auxiliary opponent modeling task. Specifically, we split the $Q$ value network into two parts: feature extractor $\mathcal{E}_{\phi}$ and decision maker $\mathcal{D}_{\psi_{a}}$. The parameters of the $Q$ value network $Q_{\theta^i}$ for agent $i$ are sequentially divided into $\phi^{i}$ and $\psi_{a}^{i}$, i.e., $\theta^{i}=(\phi^{i}, \psi_{a}^{i})$. The auxiliary opponent modeling task shares a common feature extractor $\mathcal{E}_{\phi^{i}}$ with the value network. We can update the parameters of $\mathcal{E}_{\phi^{i}}$ during execution using gradients from the auxiliary opponent modeling task, such that $\pi_{\theta^i}$ can generalize to different opponents. The supervised prediction head and its specific parameters are $\mathcal{D}_{\psi_{s}^{i}}$ with $\psi_{s}^{i}$. The details of our network architecture are shown in Fig. \ref{GRSP-AR}.

During training, the agent $i$ can collect a set of trajectory transitions $\{(o_{t}^{i}, o_{t+1}^{i}, \mathbf{a}_{t}^{-i})\}_{t=0}^{T}$ where $\mathbf{a}_{t}^{-i}$ indicates the joint actions of other agents except $i$ at time step $t$. We use the embeddings of agent $i$'s observations $o_{t}^{i}$ and $o_{t+1}^{i}$ to predict the joint actions $\mathbf{a}_{t}^{-i}$, i.e., the $\mathcal{D}_{\psi_{s}^{i}}$ is a multi-head neural network whose outputs are multiple soft-max distributions over the discrete action space or predicted continuous actions of each other agent, and the objective function of the auxiliary opponent modeling task can be formulated as
\begin{align}
&\mathcal{L}\left(o_{t}^{i}, o_{t+1}^{i},\mathbf{a}_{t}^{-i};\phi^{i}, \psi^{i}_{s}\right)\\
&=\frac{1}{N-1}\sum_{j=1,j\neq i}^{N}\ell\left({a}_{t}^{j}, \mathcal{D}_{\psi_{s}^{i}}\left(\mathcal{E}_{\phi^{i}}\left(o_{t}^{i}\right), \mathcal{E}_{\phi^{i}}\left(o_{t+1}^{i}\right)\right)^{j}\right),
\end{align}

where $\ell(\cdot)$ is the cross-entropy loss function for discrete actions or mean squared error for continuous actions. The strategies of opponents will change constantly during the procedure of multi-agent exploration and thus various strategies will emerge. The agent can leverage them to gain some experience about how to make the best response by jointly optimizing the auxiliary opponent modeling task and quantile value distribution. The joint training problem is therefore
\begin{equation}
    \min_{\phi^{i}, \psi^{i}_{s}, \psi^{i}_{a}} \mathcal{J}\left(o_{t}^{i}, a_{t}^{i}, r_{t}^{i}, o_{t+1}^{i}; \phi^{i}, \psi^{i}_{a} \right) + \mathcal{L}\left(o_{t}^{i}, o_{t+1}^{i},\mathbf{a}_{t}^{-i}; \phi^{i}, \psi^{i}_{s}\right)
\end{equation}

\subsection{Test-Time Policy Adaptation under Different Opponents}\label{adaptation}

During testing time, we can not optimize $\mathcal{J}$ anymore since the reward is unavailable, but we assume the agent can observe actions made by its opponents during execution, then we can continue optimizing $\mathcal{L}$ to update the parameters of shared feature extractor $\mathcal{E}_{\phi}$. Learning from opponents' past behaviors at test time makes the agent adapt to different opponents efficiently. This can be formulated as 
\begin{equation}
    \min_{\phi^{i}, \psi^{i}_{s}} \mathcal{L}\left(o_{t}^{i}, o_{t+1}^{i},\mathbf{a}_{t}^{-i}; \phi^{i}, \psi^{i}_{s}\right)
\end{equation}

\section{Experimental Setup}
\label{experiment}

\subsection{Environments}

\label{envsetup}

\begin{figure*}[h]
    \centering
    \includegraphics[width=\linewidth]{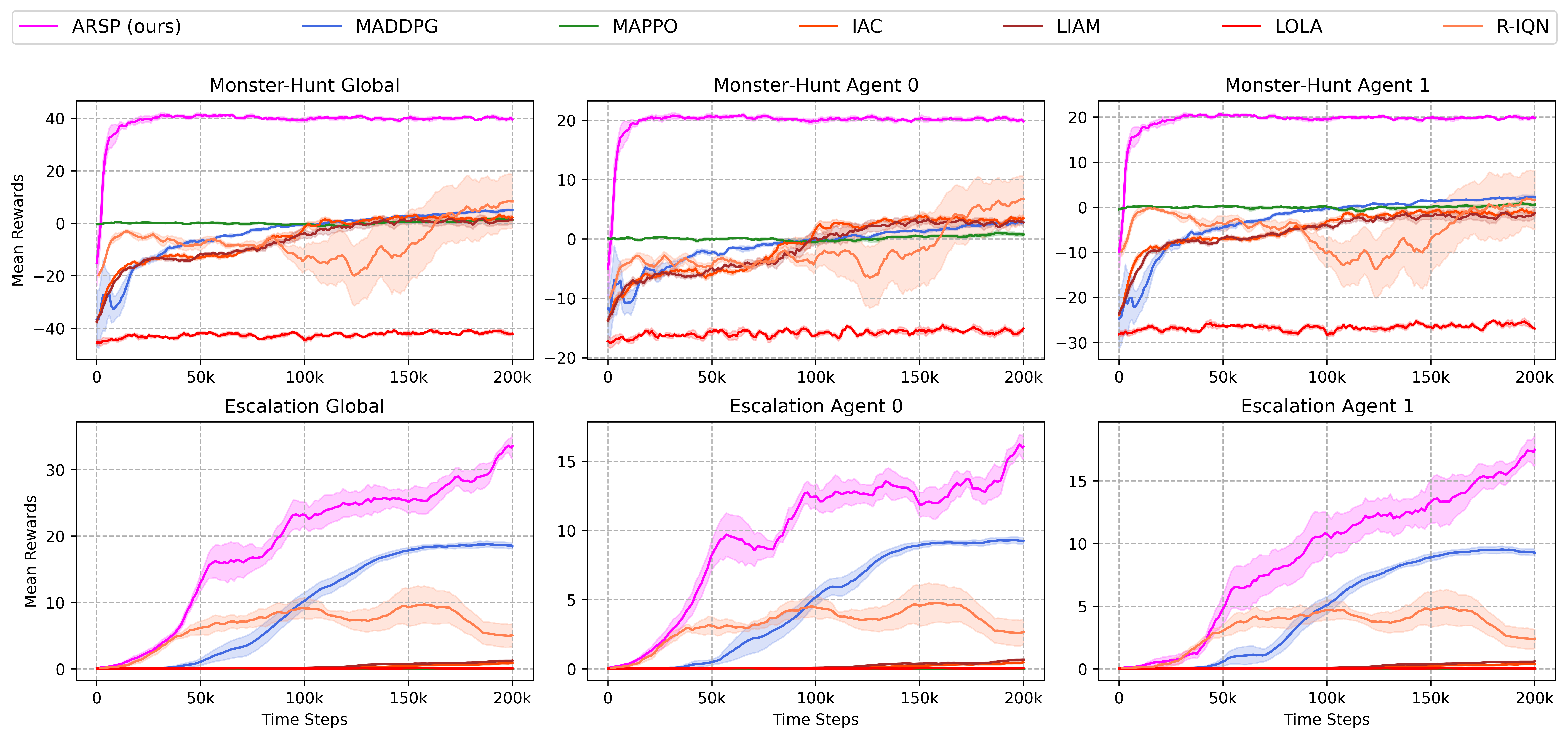}
    \caption{Mean evaluation returns for ARSP, MADDPG, MAPPO, IAC, LIAM and LOLA on Monster-Hunt and Escalation. Global rewards are summation of both agents rewards.}
    \label{fig:mhes}
\end{figure*}

s
\textbf{Iterated Prisoner’s Dilemma.} The prisoner’s dilemma is one of the most widely-studied and important general-sum games, with applications in evolutionary biology, economics, politics, sociology, and other fields. In the single-shot prisoner’s dilemma, agents can choose to cooperate (C) or defect (D) against each other, and there is only one Nash equilibrium~\cite{fudenberg1991game}, where both agents defect. The payoff of the result is presented in Table \ref{tab:ipd}. A common extension of the prisoner’s dilemma is the IPD, in which the prisoner’s dilemma is played repeatedly, with players able to observe their opponent’s past decisions. In infinitely IPD, the folk theorem~\cite{myerson1997game} shows that there are infinitely many Nash equilibria. For example, grim-trigger, in which the agent starts out cooperating and keeps cooperating as long as the other agent cooperates. But as soon as the other agent defects, the agent defect every play after that, and tit for tat (TFT), in which the agent copies the other agent’s last move. It is noteworthy that cooperative strategy generally form equilibria with each other: if both agents are playing grim-trigger, then they will keep cooperating and none of them can do any better with a different strategy.

\begin{table}[h]
\centering
\caption{Payoff Matrix for the Prisoner’s Dilemma.}
\label{tab:ipd}
\begin{tabular}{c|c|c|}
\multicolumn{1}{l|}{} & C      & D       \\ \hline
C                     & (2, 2)  & (-1, 3) \\ \hline
D                     & (3, -1) & (0, 0)   \\ \hline
\end{tabular}
\end{table}

\textbf{Iterated Stag Hunt.} Iterated Stag Hunt (ISH) is an extension of Stag Hunt matrix game similar with IPD. The one-step payoff matrix used in our experiments is shown in Table \ref{tab:ish}. There exist two NEs in ISH - keep cooperation and keep defect. Agents in both IPD and ISH can condition their actions on past history. Similar like ~\cite{lola,lu2022model}, We consider memory-1 players, i.e., the agents act based on the results of last one rounds.

\begin{table}[h]
\centering
\caption{Payoff Matrix for the Stag Hunt.}
\label{tab:ish}
\begin{tabular}{c|c|c|}
\multicolumn{1}{l|}{} & C       & D        \\ \hline
C                     & (2, 2)   & (-10, 1) \\ \hline
D                     & (1, -10) & (1, 1)    \\ \hline
\end{tabular}
\end{table}

\begin{figure}[]
    \centering
    \includegraphics[width=\linewidth]{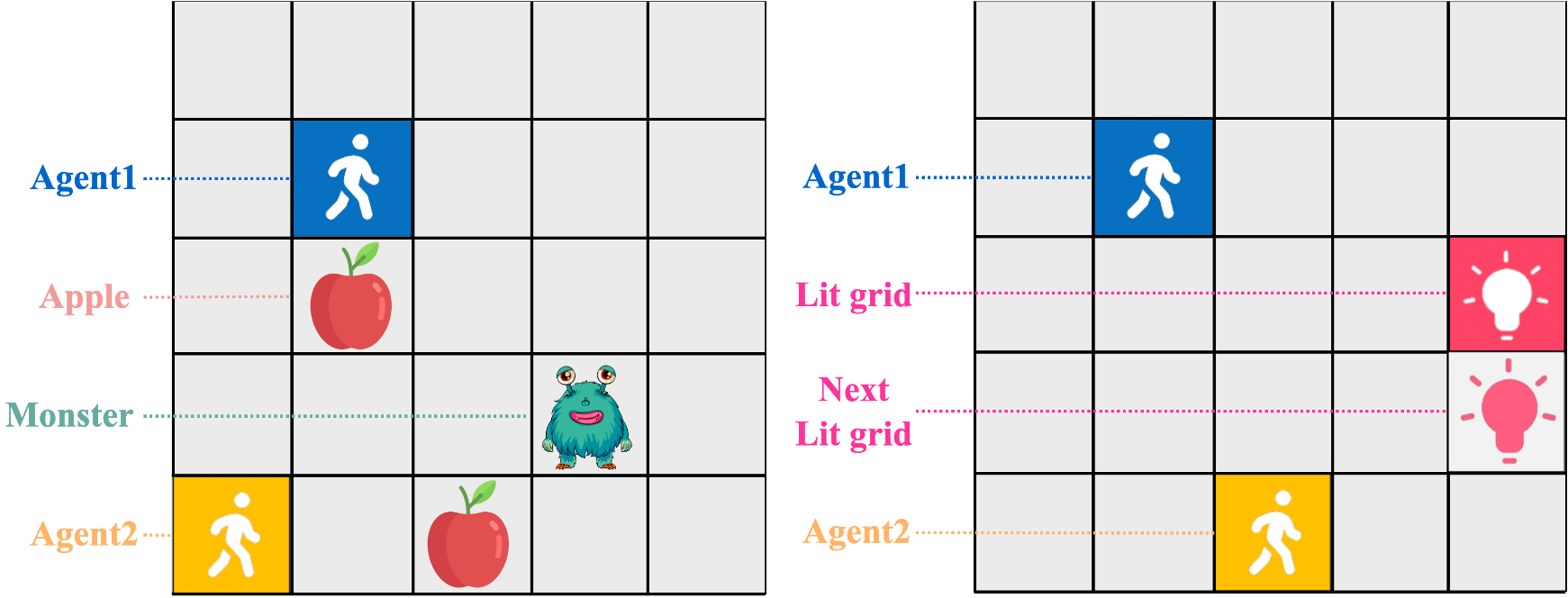}
    \caption{Monster-Hunt and Escalation}
    \label{fig:my_label}
\end{figure}

\textbf{Monster-Hunt (MH).} The Monster-Hunt~\cite{rpg} is a multi-agent grid-world environment with high dimensional state space. It is a $5\times5$ grid and consists of two agents, two apples and one monster. The apples are static while the monster keeps moving towards its closest agent. When a single agent meets the monster, it gets a penalty of -10. If two agents catch the monster together, they both get a bonus of 5. If one agent meets an apple, it gets a bonus of 2. Whenever an apple is eaten or the monster meets an agent, the entity will respawn randomly in the grid. The cooperative strategy, i.e., both agents catch the monster together, is risky since the agent will suffer a big loss if the other agent defect. The episode length is 20 in our experiment.

\textbf{Escalation.} Escalation is a $5\times5$ grid-world with sparse rewards, consisting of two agents and a static light. If both agents step on the light simultaneously, they receive a bonus of 1, and then the light moves to a random adjacent grid. If only one agent steps on the light, he gets a penalty of $1.5\times L$, where $L$ denotes the latest consecutive cooperation steps, and the light will respawn randomly. To maximize their individual payoffs, agents must coordinate to stay together and step on the light grid simultaneously. For each integer L, there is a corresponding coordination strategy where each agent follows the light for $L$ steps then simultaneously stop coordination. However, the agent will suffer more losses once the other agent defect with the increase of cooperation steps. The episode length is 30 in our experiment.

\begin{figure*}
    \centering
    \includegraphics[width=\linewidth]{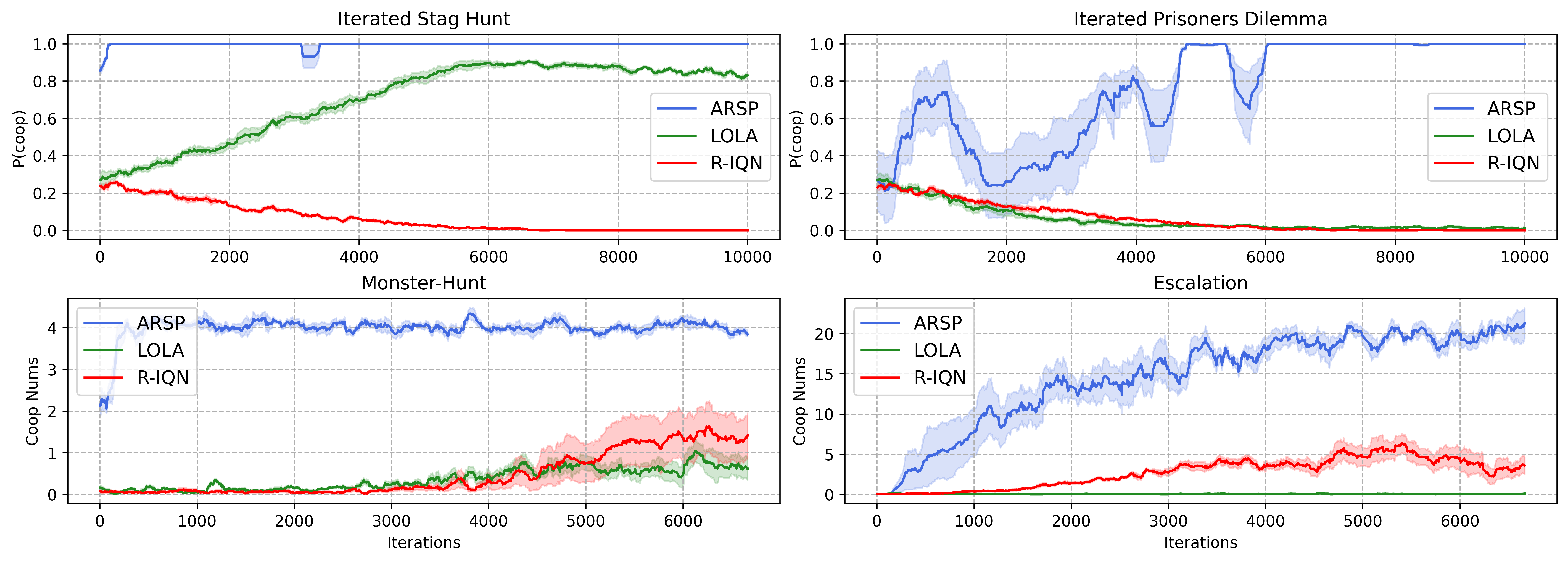}
    \caption{The probability of agents cooperate with each other in ISH and IPD (first row) during training and the number of mutual cooperation steps in one episode in Monster-Hunt and Escalation (second row) during training.}
    \label{fig:coop}
\end{figure*}

\subsection{Baseline Comparisons}

\begin{itemize}
    \item \textbf{Independent Actor-Critic (IAC): } IAC is a naive decentralized policy gradient method. Each agent learns policy and value networks based on its local observations and treat other agents as part of the environment.
    
    \item \textbf{Local Information Agent Modelling (LIAM):} LIAM~\cite{liam} learns latent representations of the opponent from the ego agent's local information using encoder-decoder architecture. The opponent's observations and actions are utilized as reconstruction targets for the decoder, and the learned latent representation conditions the policy of the ego agent in addition to its local observation. The policy and model are optimized based on A2C algorithm.
    
    \item \textbf{Learning with Opponent Learning Awareness (LOLA): }
    LOLA~\cite{lola} assumes that other agents are naive learners and considers the learning processes of other agents. LOLA takes a gradient through the opponent’s update function to shape the opponent. In the self-play setting, LOLA is one of the first methods to discover the tit-for-tat (TFT) strategy in the IPD.

    \item \textbf{Multi-Agent Deep Deterministic Policy Gradient (MADDPG): }MADDPG~\cite{maddpg} is an extension of DDPG~\cite{ddpg} methods in MARL where the critic is augmented with extra information about the policies of other agents, and is applicable to mixed cooperative-competitive environments.
    
    \item \textbf{Multi-Agent Proximal Policy Optimization (MAPPO): }MAPPO~\cite{yu2021surprising} is a variant of PPO which is specialized for multi-agent settings. MAPPO learns a centralized critic and adopts a collections of tricks, e.g., input normalization and layer normalization to improve agent's performance.
    
    \item \textbf{Risk-Sensitive Implicit Quantile Networks (R-IQN): }R-IQN~\cite{iqn} is a dencentralized risk-sensitive distributional reinforcement learning method. It learns  the quantile values for sampled quantile fractions with an implicit quantile value network (IQN) that maps from quantile fractions to quantile values. By sampling quantile fractions $\tau$ from different distortion risk measures instead of uniform distribution, agents can be risk-seeking or risk-averse. We use WT as the distortion risk measure to construct risk-seeking agents and compare them with our ARSP agents.

\end{itemize}

\section{Results}
\label{eval}

In this subsection, we evaluate all methods on four multi-agent environments and use 5 different random seeds to evaluate each method. We pause training every 50 episodes and run 30 independent episodes to evaluate the average performance of each method.

\subsection{Iterated Games}

Fig. \ref{fig:img} shows the average global rewards, i.e., the summation of all agents' average returns, of all methods evaluated during training in ISH and IPD. The shadowed part represents a $95\%$ confidence interval. The average global rewards equal to 40 means that both agents' have discovered coordination strategy, i.e., cooperating at each time step. ARSP vastly outperforms all other learning methods in the IPD and ISH. Notably, it is the only algorithm to achieve the optimal coordination strategy stably in a sample efficient way. LOLA outperforms other baseline methods except ARSP in ISH and IPD. However, LOLA agents are unable to achieve coordination stably. R-IQN is a risk-seeking method similar to ARSP but failed to discover coordination strategies. We suspect that this happens because R-IQN implicitly achieves risk-seeking by sampling quantile fractions from the distorted distribution instead of uniform, whereas ARSP explicitly constructs risk-seeking exploration bonus to encourage agents to discover coordination strategies more efficiently. The first row of Fig. \ref{fig:coop}  shows the probability of agents cooperate with each other in ISH and IPD during training. ARSP agents can converge to coordination strategies efficiently in probability one. To our best knowledge, it is the first method to achieve this result.


\begin{figure*}[h]
    \centering
    \includegraphics[width=\linewidth]{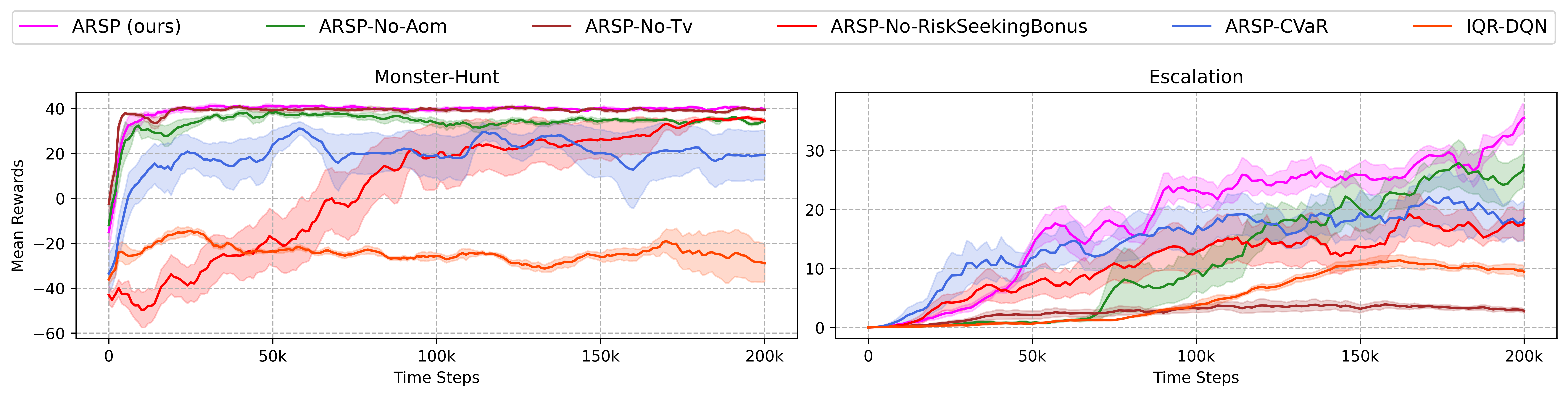}
    \caption{Mean evaluation return of ARSP compared with other ablation methods in two grid-world multi-agent environments.}
    \label{fig:ablation}
\end{figure*}

\subsection{Grid-Worlds}

We further show the effectiveness of ARSP in two high dimensional grid-world games - MH and Escalation~\cite{rpg}. Both of them have multiple NEs with different payoff.

The first row of Fig. \ref{fig:mhes} is the evaluation results of ARSP and other baselines in MH. Global rewards are the summation of all agents' returns in the environment. In MH, ARSP agents can rapidly discover the high payoff NE where two agents stay together and wait for the monster. Notably, ARSP agents can stably converge to the coordination strategy in less than 25k time steps, which shows the superiority of ARSP in sample efficiency. However, most other baseline agents can only converge to some guaranteed but lower payoff NEs, i.e., avoid stag and eat apples alone. LOLA do not achieve significant results in MH, and similar results also appear in ~\cite{foerster2018dice}. We suspect that LOLA can not scale to high dimensional settings. R-IQN performs better than other baselines in MH and the second row of Fig \ref{fig:coop} shows that R-IQN agents can achieve one or two co-operations in one episode, demonstrating that the risk-seeking policy plays an important role in discovering coordination strategies. Similar results are also achieved in the Escalation environment, where ARSP significantly outperforms other baselines both in asymptotic performance and sample efficiency. Although with the increase of cooperation times, the punishment suffered by agents for betrayal is also increasing in the Escalation, agents can still achieve stable cooperation shown in Fig \ref{fig:coop}.


\subsection{Adaptation Study}
\label{generalization}

This subsection investigates the ability of the pre-trained ARSP agent to adapt to different opponents during execution. The cooperative opponents are trained by ARSP method because ARSP is the only method to produce cooperative strategies, and the non-cooperative opponents are MADDPG agents. During evaluation, random seeds of four environments are different from that during training, and hyperparameters of the ARSP, e.g., risk-seeking level $\lambda$, are same and fixed between different opponents. Furthermore, the pre-trained agent can no longer use environment rewards to update its policy and it must utilize the auxiliary opponent modeling task to adapt to different opponents. The network details and hyperparameters can be found in the Appendix.


\begin{table}[h]
\caption{Mean evaluation return of ARSP with and without auxiliary opponent modeling task on four multi-agent environments.}
\label{tab:gs}
\begin{tabular}{|c|cc|cc|}
\hline
\multicolumn{1}{|l|}{\multirow{2}{*}{}} & \multicolumn{2}{c|}{ARSP-No-Aom}                         & \multicolumn{2}{c|}{ARSP}           \\
\multicolumn{1}{|l|}{}                  & \multicolumn{1}{c|}{CC}    & CD                          & \multicolumn{1}{c|}{CC}    & CD     \\ \hline
ISH                                     & \multicolumn{1}{c|}{$20$}    & $-100$                        & \multicolumn{1}{c|}{$\mathbf{20}$}    & $\mathbf{0.65}$   \\ \hline
IPD                                     & \multicolumn{1}{c|}{$20$}    & $-5$                          & \multicolumn{1}{c|}{$\mathbf{20}$}    & $\mathbf{-1.08}$  \\ \hline
M-H                                     & \multicolumn{1}{c|}{$20.62$} & \multicolumn{1}{l|}{$-15.03$} & \multicolumn{1}{c|}{$\mathbf{21.36}$} & $\mathbf{-12.07}$ \\ \hline
Escalation                              & \multicolumn{1}{c|}{$9.45$}  & \multicolumn{1}{l|}{$-0.545$} & \multicolumn{1}{c|}{$\mathbf{11.3}$}  & $\mathbf{0.175}$  \\ \hline
\end{tabular}
\end{table}

Table \ref{tab:gs} shows the mean evaluation return of the ARSP agent with and without auxiliary opponent modeling (Aom) task on four multi-agent environments when interacting with different opponents. All returns are averaged on 100 episodes. ARSP-No-Aom means that the agent is trained without Aom task. CC indicates the opponent is a cooperative agent while CD means the opponent will defect. In ISH and IPD, the ARSP agent chooses to cooperate at the first time step. After observing its opponent's decision, it will choose to keep cooperating or defecting, depending on its opponent's type. So the ARSP agent can avoid being exploited by its opponent and gets guaranteed rewards, while ARSP-No-Aom can not, as shown in Table \ref{tab:gs}. Similar results also appear in M-H and Escalation. Furthermore, when interacting with cooperative opponents, ARSP agent can adapt to their policies and receive higher individual rewards than ARSP-No-Aom, e.g., 21.36 in M-H and 11.3 in Escalation. The experiment results also demonstrate that policies learned independently can overfit other agents' policies during training, and our Aom method provides a way to tackle this problem.

\subsection{Ablations}
\label{ablations}

In this subsection, we perform an ablation study to examine the components of ARSP to better understand our method. ARSP is based on QR-DQN and has three components: the risk-seeking exploration bonus, the left truncated variance (Tv) and the auxiliary opponent modeling task (Aom). We design and evaluate six different ablations of ARSP in two grid-world environments, as show in Fig. \ref{fig:ablation}. The evaluation return of ARSP-No-Aom which we ablate the Aom module and retain all other features of our method is a little lower than that of ARSP but has a much higher variance, indicating that learning from opponent's behaviors can stable training and improve performance. Moreover, the ARSP-No-Aom is a completely decentralized method whose training without any opponent information, and the ablation results of ARSP-No-Aom indicate that our risk-seeking bonus is the determining factor for agents to achieve coordination strategies in our experiments. We observe that ablating the left truncated variance module leads to a lower return than ARSP in the Escalation but no difference in the Monster-Hunt. Furthermore, ablating the risk-seeking bonus increases the training variance, leads to slower convergence and worse performance. It is noteworthy that the Escalation is a sparse reward and hard-exploration multi-agent environment because two decentralized agents can not get any reward until they navigate to and step on the light simultaneously and constantly. These two ablations indicate that the exploration ability of left truncated variance is important for our method and the risk-seeking bonus can encourage agents to coordinate with each other stably and converge to high-risky cooperation strategies efficiently. We also implement our risk-seeking bonus with CVaR instead of WT, and the results are shown as ARSP-CVaR. The ARSP-CVaR performs worse than our method and has a higher training variance. Finally, we ablate all components of the ARSP  and use $\epsilon$-greedy policy for exploration which leads to the IQR-DQN algorithm. As shown in Fig. \ref{fig:ablation}, IQR-DQN can not learn effective policies both in the Monster-Hunt and the Escalation.

\section{Conclusion \& Future Work}
\label{discu}

In this paper, we presented Adaptable Risk-Sensitive Policy (ARSP), a novel adaptable risk-sensitive reinforcement learning method for multi-agent general-sum games where win-win outcomes are only achieved through coordination. By estimating the risk-seeking exploration bonus, ARSP agents can efficiently discover risky coordination strategies and converge to high payoff NEs stably. To avoid overfitting training opponents and learn adaptable policies, we propose the auxiliary opponent modeling task, which leverages the opponent’s history behaviors to infer its strategies and alter the ego agent’s policy dynamically during execution, thus adapting to different opponents.

More specifically, ARSP agents can discover the optimal coordination strategy in both ISH and IPD and converge to it in probability one. To the best of our knowledge, it is the first method to achieve this result. Furthermore, ARSP can also scale to more complex, high-dimensional multi-agent games and achieve similar results in Monster-Hunt and Escalation. We also show that the ARSP agent can efficiently adapt to its opponent's policy, avoid being exploited by non-cooperative opponent and further improve its coordination performance for cooperative opponent. 

The risk-seeking bonus in ARSP is estimated using WT distorted expectation and its risk-sensitive level is a hyperparameter. Developing the method that can adjust agent' risk-sensitive level dynamically by utilizing its observation, rewards, or opponents' information is the direction of our future work.

\bibliographystyle{named}
\bibliography{ijcai22}

\end{document}